\documentclass[runningheads]{llncs}
\usepackage[T1]{fontenc}
\usepackage{graphicx}
\usepackage{wrapfig}
\usepackage{multicol}
\usepackage{multirow}

\begin{document}
\title{A Design of A Simple Yet Effective Exercise Recommendation System in K-12 Online Learning}
\titlerunning{A Design of A Simple Yet Effective Exercise Recommendation System}
% If the paper title is too long for the running head, you can set
% an abbreviated paper title here
%

\author{Shuyan Huang\inst{1}\orcidID{0000-0003-0217-7494} \and
Qiongqiong Liu\inst{1}\orcidID{0000-0001-7842-2447} \and
Jiahao Chen\inst{1}\orcidID{0000-0001-6095-1041} \and
Xiangen Hu\inst{2}\orcidID{1111-2222-3333-4444} \and
Zitao Liu\inst{1}\orcidID{0000-0003-0491-307X}\thanks{The corresponding author: Zitao Liu} \and
Weiqi Luo\inst{3}\orcidID{0000-0001-5605-7397}}
\authorrunning{S. Huang et al.}
% First names are abbreviated in the running head.
% If there are more than two authors, 'et al.' is used.
%
\institute{TAL Education Group, Beijing, China\\
\email{\{huangshuyan, liuqiongqiong1, chenjiahao, liuzitao\}@tal.com}\and
The University of Memphis, Memphis, TN USA\\
\email{xhu@memphis.edu}\and
Guangdong Institute of Smart Education, Jinan University, Guangzhou, China\\
\email{lwq@jnu.edu}}
\maketitle              % typeset the header of the contribution
\begin{abstract}
We propose a simple but effective method to recommend exercises with high quality and diversity for students. Our method is made up of three key components: (1) candidate generation module; (2) diversity-promoting module; and (3) scope restriction module. The proposed method improves the overall recommendation performance in terms of recall, and increases the diversity of the recommended candidates by 0.81\% compared to the baselines.

\end{abstract}

\section{Introduction}
\label{sec:intro}

The recommender systems (RSs) are widely adopted in internet applications to provide relevant items according to users' preference history \cite{schafer2001commerce,barria2021explainable,liu2019recommender,yan2018practical}. In the online learning scenario, users are learners and items are exercises. As a key component in online learning, exercise recommendation systems provide personalized exercises to every student according to his or her individual requirements \cite{wang2019adaptive,mohseni2019pique,huang2019exploring}. Despite the promising results achieved by the existing exercise recommendation methods \cite{wan2018learning,pang2017collaborative,xu2021personalized,yang2014forum,chen2021educational}, building an effective exercise RS still presents many challenges that come from special characteristics of real-life student learning scenarios. First, classic RS approaches are likely to recommend lexically similar exercises while a desired RS is supposed to generate exercises with different content but the same knowledge concepts (KCs) and help students avoid rote memorization. Second, compared to the data scale in internet companies, student-exercise interaction datasets are much smaller and even more sparse. Third, the exercise RSs accompany students for the entire semester and the recommended results should follow the learning progress of the in-school syllabus. Hence, we propose a simple yet effective exercise recommendation system to recommend KC-relevant exercises within the scope of the in-school syllabus. 
% Therefore, we propose a simple yet effective exercise recommendation system that is able to recommend KC relevant exercises in real-time within the scope of in-school syllabus. 

% Our RS focuses on capturing the similarities underlying each exercises based on their KCs instead of lexical content. Meanwhile, our approach is able to take teachers' in-class materials, i.e., slides, into considerations and restrict the recommended results along with the learner's in-school learning progress. We evaluate our RS on a real-world educational dataset and demonstrate that our approach significantly improves the performance of the generated recommendations.

\section{System Design}
\label{sec:method}
\begin{figure}
    \centering
    \includegraphics[width=1\textwidth]{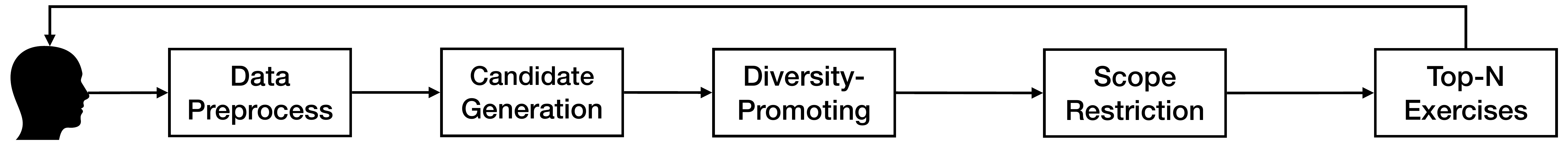}
    \vspace{-0.3cm}
    \caption{The overview of our exercise recommendation system.}
    \label{fig:overview}
    \vspace{-0.4cm}
\end{figure}

The entire exercise recommendation workflow is illustrated in Figure \ref{fig:overview}.

% Given a learner, our RS first fetches the learner's study history, which may include the past incorrectly answered problems, the relevant KCs taught in classes, etc. After obtaining and pre-processing such historical information from the learner, they are feed into our candidate generation module. Our candidate generation is implemented via an enhanced random walk algorithm on a tripartite graph made of exercises, KCs and class materials, as shown in Figure \ref{fig:overview}(a). Then all the candidates are passed through a diversity-promoting module that removes recommended exercises with high lexically similarity, shown in Figure \ref{fig:overview}(b). Finally, the remaining results are processed by a scope restriction module, which enforces the recommended results aligning with the current learning progress of a student in school, illustrated in Figure \ref{fig:overview}(c). We provide more details about the above three key components as follows.

\vspace{-0.3cm}

\noindent \textbf{Candidate Generation Module} We conduct the candidate generation by using an enhanced random walk algorithm on an educational tripartite graph that is composed of a set of KCs, exercises and class materials. By incorporating KCs and class materials in the graph, we are able to capture the underlying knowledge similarities of exercises while focusing on the in-class teaching content. For each student, we create a query set $Q$ that may contain the most recently incorrectly answered exercises as well as exercises practiced in the class materials. Then, we generate recommendations using the enhanced random walk algorithm given the query set. Let $E_1$ and $E_2$ be the edges between KCs and exercises and edges between exercises and class materials. Each random walk produces a sequence of steps. Each step is composed of three operations. First, given the current exercise \emph{q}, i.e., $q \in Q$, we select an edge $e$ from $E$, i.e., $E = E_1 \cup E_2$. If $e$ comes from $E_1$, it means $q$ is connected with a KC $k$. Otherwise, $e$ connects $q$ with a class material $m$. Then, we select exercise $e'$ by sampling an edge $e'$ from either $E_1(k)$ that connects $k$ and $e'$ or $E_2(m)$ that connects $m$ and $e'$. Third, the current exercise is updated to $e'$ and the step repeats. We use $E_1(k)$ and $E_2(m)$ to denote exercises connect to KC $k$ and class material $m$ respectively. The length of each short random walk and the total number of steps across all such short random walks are determined by two hyper-parameters $\alpha$ and $T$ respectively. We record the visit count for each candidate exercise $q$ in each step and maintain a counter $V$ that maps the candidate to the visit count. Finally, the $N$ exercises with the highest visit count are selected as the candidate recommendations.

\noindent \textbf{Diversity-Promoting Module} In this module, we remove candidates with high lexical similarities. First, we further pre-train the widely used large-scale language model, i.e., $RoBERTa$, using the exercise data \cite{liu2019roberta}. Then we obtain the exercise semantic representations by averaging all the word-level representations of its textual content. After that, similarities between all the exercises are calculated via their exercise embedding and the recommended candidates with high similarities, i.e., over a certain threshold, are filtered out. In this way, we provide a set of diverse exercises to help students avoid rote memorization.

\noindent \textbf{Scope Restriction Module} In K-12 education, the recommended exercises should align with the in-school learning progress of the students. To achieve this property, we first roughly sort all the KCs in sequential order according to the class syllabus arrangement in each semester. Then, we only select and recommend candidates whose KCs are never beyond the scope of KCs covered by the up-to-date course progress.

\section{Experiments}
\label{sec:experiment}
We conduct our method on a real-world K-12 online learning dataset with 1679 KCs, 1,153,690 exercises and 135,637 class materials. These class materials are created by teaching professionals from a third-party online educational platform\footnote{https://www.xesvip.com/}. We randomly withhold test data of 244 class materials, which contain 494 KCs and 8,340 exercises. Our exercise recommendation evaluation task is \emph{finding a set of exercises which are relevant to the set of KCs from the student's incorrectly answer exercises of the handcrafted class material.}
% Our exercise recommendation evaluation task is \emph{we aim to find a set of relevant exercises, given the set of KCs of each handcrafted class material which corresponds with the exercises that the students need to intensive practice.} 
Similar to \cite{eksombatchai2018pixie}, the recommendation performance is evaluated by (1) recall, which measures the percentages of recommended exercises that the handcrafted class material contains as well; and (2) Distinct-2 , which measures the diversity of selected exercises by computing the ratios of the number of different bi-grams to the total number of bi-grams \cite{li2016diversity}. We compute recall scores from both micro and macro perspectives \cite{van2013macro}. We set the $\alpha$ is 0.04 and the total steps $T$ of a random walk procedure is 100,000. For each KC, we select the top $N \in \{10,25,100\}$ recommended exercise candidates. 

\vspace{-0.3cm}
\begin{table}
\caption{\label{tab:results} Results (\%) of different recommendation methods on the online learning dataset. TG, DP, and SR denote the tripartite graph based candidate generation module, the diversity-promoting module and the scope restriction module respectively.}
\centering
\scalebox{0.7}{
\begin{tabular}{|l|c|c|c|c|c|c|c|} 
\hline
\multicolumn{1}{|c|}{\multirow{2}{*}{Model~}} & \multicolumn{2}{c|}{Top-10} & \multicolumn{2}{c|}{Top-25} & \multicolumn{2}{c|}{Top-100} & \multicolumn{1}{l|}{\multirow{2}{*}{Distinct-2}}  \\ 
\cline{2-7}
\multicolumn{1}{|c|}{}                        & Macro Recall & Micro Recall & Macro Recall & Micro Recall & Macro Recall & Micro Recall  & \multicolumn{1}{l|}{}                             \\ 
\hline
\hline
TG                                            & 13.52        & 9.63         & 25.73        & 19.48        & 44.70        & 36.22         & 13.05                                             \\ 
\hline
TG+DP                                         & 13.52        & 9.64         & 25.69        & 19.48        & 44.41        & 36.06         & 13.21                                             \\ 
\hline
TG+SR                                         & 13.75        & 9.80         & \textbf{26.25}        & \textbf{19.90}        & \textbf{45.71}        & \textbf{36.93}         & 13.69                                             \\ 
\hline
TG+DP+SR                                      & \textbf{13.80}        & \textbf{9.82}         & 26.20        & 19.86        & 45.42        & 36.80         &  \textbf{13.86}                                             \\ 
\hline
\end{tabular}}
\vspace{-0.4cm}
\end{table}

The recommendation results are shown in Table \ref{tab:results}. From Table \ref{tab:results}, we make the following observations: First, when comparing TG+DP+SR to TG+DP, we can see that SR in general improves the overall recommendation performance. We believe this is because the SR module is able to remove the out-of-scope recommended candidates generated from the random walk based candidate generation module. Second, comparing TG and TG+DP, TG+SR and TG+DP+SR, we can see, the DP module increases the diversity of the recommended exercises up to 0.17\%.

% First, by comparing the results of all the models in three datasets, eg. TG in \emph{D1} vs. TG in \emph{D2} and TG+DP+SR in \emph{D2} vs. TG+DP+SR in \emph{D3}, we can see that the recall decreases with the increase of the number of KCs in the class materials. This indicates that the recommended results for a large set of KCs are more diverse compared to recommendations generated from a smaller KC set. 

% Second, in all three datasets, SR improves the overall recommendation performance. For example, the macro recall of Top 100 recommendation results from TG+DP+SR on \emph{D1} is 1.58\% higher compared to recall of TG+DP. This is because there are likely recommended exercises beyond the scope of the class from the random walk based candidate generation module and remove such out-of-scope exercises will improve the recommendation performance.

% in Figure \ref{fig:diversity}, we utilize the Distinct-2 to measure the diversity of selected exercises by computing the ratios of the number of different bi-grams to the total number of bi-grams \cite{li2016diversity}. Comparing TG and TG+DP in all three datasets, the DP module increases the diversity of the recommended exercises. The diversity is decreasing with the increasing of the thresholds of similarity, because the influence for the final results of DP decreases with larger threshold. 

% \begin{wrapfigure}{l}{6cm}
% \vspace{-0.85cm}
% \includegraphics[width=0.5\textwidth]{diversity.pdf}\vspace{-0.5cm}
% \caption{Diversity of different thresholds} 
% \label{fig:diversity}
% \vspace{-0.7cm}
% \end{wrapfigure}

\section{Conclusion}
\label{sec:conclusion}
In this work, we propose a simple but effective method utilizing exercises, KCs, and class materials to recommend exercises for students. Our recommendation framework relies on the candidate generation module, the diversity-promoting module and the scope restriction module. We evaluate our approach on an offline educational data set and the results show that the proposed method is able to give accurate and more diverse recommended exercises. 

\subsubsection{Acknowledgements} This work was supported in part by National Key R\&D Program of China, under Grant No. 2020AAA0104500; in part by Beijing Nova Program (Z201100006820068) from Beijing Municipal Science \& Technology Commission and in part by NFSC under Grant No. 61877029.

%
% ---- Bibliography ----
%
% BibTeX users should specify bibliography style 'splncs04'.
% References will then be sorted and formatted in the correct style.

\bibliographystyle{splncs04}
\bibliography{aied2022}

\end{document}